\documentclass{pnastwo}

\newcommand{\aj} {Astron.~J.}
\newcommand{\aap}{Astron.~Astrophys.}
\newcommand{\apj}{Astrophys.~J.}
\newcommand{\apjs}{Astrophys.~J.~Supp.}
\newcommand{\apjl}{Astrophys.~J.~Let.}

\newcommand{\nat}{Nature}
\newcommand{\icarus}{Icarus}
\newcommand{\mnras}{Mon.~Not.~R.~Astron.~Soc.}

\renewcommand{\thetable}{S\arabic{table}}

\url{www.pnas.org/cgi/doi/}
\copyrightyear{2015}
\issuedate{Issue Date}
\volume{Volume}
\issuenumber{Issue Number}
\begin{document}

\title{Growing the Terrestrial Planets from the Gradual Accumulation of Sub-Meter Sized Objects}

\author{Harold F$.$ Levison\affil{1}{Southwest Research Institute and NASA Solar System Exploration Research Virtual Institute, 1050 Walnut St, Suite 300, Boulder, Colorado 80302, USA}\footnote{
	To whom correspondence should be addressed. Email: {hal@boulder.swri.edu}},
Katherine A$.$ Kretke\affil{1}{},
Kevin Walsh\affil{1}{}, \and
 William Bottke\affil{1}{}}

\contributor{Submitted to Proceedings of the National Academy of Sciences
of the United States of America}

\significancetext{The fact that Mars is so much smaller than both the
  Earth and Venus has been a long standing puzzle of terrestrial
  planet formation.  Here we show that a new mode of planet formation
  known as `Viscous Stirred Pebble Accretion', which as recently been
  shown to produce the giant planets, also naturally explains the
  small size of Mars and the low mass of the asteroid belt.  Thus
  there is a unified model that can be used to explain the all of the
  basic properties of the our Solar System.}

\maketitle

\begin{article}
\begin{abstract}
Building the terrestrial planets has been a challenge for planet formation
models.  In particular, classical theories have been unable to reproduce the
small mass of Mars and instead predict that a planet near 1.5 AU should roughly
be the same mass as the Earth.  Recently, a new model called \emph{Viscous
Stirred Pebble Accretion} (VSPA) has been developed that can explain the
formation of the gas giants.  This model envisions that the cores of the giant
planets formed from 100 to 1000 km bodies that directly accreted a population
of pebbles --- sub-meter sized objects that slowly grew in the protoplanetary
disk.  Here we apply this model to the terrestrial planet region and find that
it can reproduce the basic structure of the inner Solar System, including a
small Mars and a low-mass asteroid belt.  Our models show that for an initial
population of planetesimals with sizes similar to those of the main belt
asteroids, VSPA becomes inefficient beyond $\sim\!$1.5 AU.  As a result, Mars's
growth is stunted and nothing large in the asteroid belt can accumulate.
\end{abstract}


\abbreviations{VSPA, viscously stirred pebble accretion}

\dropcap{C}lassical models of terrestrial planet formation have a
problem, the same models that produce reasonable Earth and Venus
analogs tend to produce Mars analogs that are far too large
\cite{Raymond.etal.2009}.  The only existing proposed
explanations for the small mass of Mars based on classical modes
  of growth require a severe depletion solids beyond $1\,$AU
  \cite{Hansen.2009} involving either not-well-understood nebular
  processes \cite{Izidoro.etal.2014} or a complicated and dramatic
migration of the giant planets \cite{Walsh.etal.2011} to solve this
problem.  Recently, however, it has been shown that a new mode of
planet formation known as \emph{Viscous Stirred Pebble Accretion}
(VSPA) can successfully explain the formation of the giant planets
\cite{Lambrechts.Johansen.2012,Levison.etal.2015a}.  Here it is our
hypothesis that Mars's mass may simply be another manifestation of
VSPA.  To understand how, we need to describe the process.

\section{Review of Pebble Accretion}
After the formation of the protoplanetary disk, dust particles, which
are suspended in the gas, slowly collide and grow because of
electrostatic forces.  Once particles become large enough so that
their Stokes numbers ($\tau \equiv t_s \Omega_K$, where $t_s$ is the
stopping time due to aerodynamic drag and $\Omega_K$ is the orbital
frequency) are between $\sim\!10^{-3}$ and 1, depending on the model,
these so-called `pebbles' can be concentrated by aerodynamic processes
\cite{Cuzzi.etal.2001, Youdin.Goodman.2005, Cuzzi.etal.2008,
  Johansen.etal.2007}. Under the appropriate physical conditions
  (which might not have been satisfied everywhere in the disk), these
concentrations become dense enough that they become gravitationally
unstable and thus collapse to form planetesimals
\cite{Goldreich.Ward.1973} with radii between $\sim\!50$ and
$\sim\!1000\,$km \cite{Johansen.etal.2007,Youdin.2011a}.  This process
can occur very quickly --- on the order of the local orbital period.

Recent research shows that planetesimals embedded in a population of
pebbles can grow rapidly via a newly discovered accretion mechanism
that is aided by aerodynamic drag on the pebbles themselves
\cite{Ormel.Klahr.2010,Lambrechts.Johansen.2012,Morbidelli.Nesvorny.2012}.
In particular, if a pebble's aerodynamic stopping time is less than or
comparable to the time for it to encounter a growing body (hereafter
known as an {\it embryo}) then it is decelerated with respect to the
embryo and becomes gravitationally bound.  After capture, the pebble
spirals towards the embryo due to aerodynamic drag and is accreted.
The accretional cross section for this situation is
\begin{equation}
  \sigma_{\rm peb} \equiv \pi \frac{4 G M_e t_s}{v_{rel}} \exp^{-\xi},
	\label{eq:R_C}
\end{equation}
where $\xi = 2\left[t_s v_{rel}^3/(4G M_e)\right]^{0.65}$, $M_e$ is
the mass of the embryo, and $v_{rel}$ is the relative velocity between
the pebble and embryo \cite{Ormel.Klahr.2010}.  For the growing
planets, $\sigma_{\rm peb}$ can be orders of magnitude larger than the
physical cross section alone.  Full $N$-body simulations
\cite{Levison.etal.2015a} show that as long as pebbles form
continuously over a long enough time period such that embryos have
time to gravitationally stir each other, this process can form the
observed gas giant planets before the gas disk dissipates.

\begin{figure}
	\centerline{\includegraphics[width=.5\textwidth]{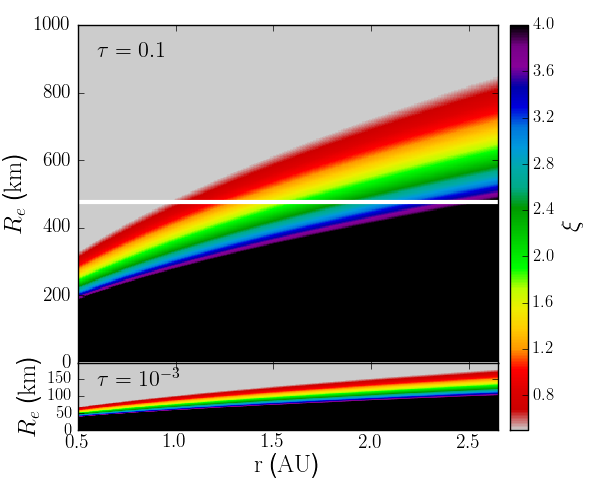}}
	\caption{The value of $\xi$ as a function of heliocentric
          distance ($r$) and embryo radius ($R_e$) in our fiducial
          protoplanetary disk, assuming that the embryos have
            circular (Keplerian) orbits and the pebbles are on orbits
            as determined by aerodynamic drag. The top and bottom
          panels are for $\tau=0.1$ and $10^{-3}$, respectively.  As
          $\xi$ in general increases with heliocentric distance for an
          object of a given size, objects that can grow in the inner
          regions cannot grow farther from the Sun.  For reference,
          the white horizontal line in the top panel corresponds to
          the radius of (1) Ceres.}  \label{fig:XI} \end{figure}

Our hypothesis that this process can also explain Mars's small size
and the low mass of the asteroid belt is based on the $e^{-\xi}$ term
in Eq$.$~(\ref{eq:R_C}), which says that pebble accretion becomes
exponentially less efficient for small embryos because the encounter
times for these objects ($4G M_e/v_{rel}^3$) becomes short compared to
$t_s$.  As a result, aerodynamic drag does not have time to change the
trajectory of the pebbles, so they are unlikely to be accreted.
Eq$.$~(\ref{eq:R_C}) therefore predicts a sharp cutoff between small
embryos, which cannot grow, and larger objects, which can.  In
addition, because $t_s$ is a function of location in the disk, this
cutoff also varies with location.  Fig$.$~\ref{fig:XI} shows the value
of $\xi$ in our fiducial disk (which is described in the {\it
    Methods} section) for two values of $\tau$ that are consistent
with the requirements of the two competing models of planetesimal
formation.  In particular, the top panel employs $\tau=0.1$ pebbles
that are required for the so-called {\it streaming instability}
\cite{Youdin.Goodman.2005, Johansen.etal.2007, Bai.Stone.2010}, while
the bottom panel uses the $\tau=10^{-3}$ pebbles needed by the
turbulent concentration models of Ref.~\cite{Cuzzi.etal.2001,
  Cuzzi.etal.2008}.  As $\xi$ in general increases with heliocentric
distance for an embryo of a given size, embryos that can grow in the
inner regions cannot grow farther from the Sun. For example, if
pebbles have $\tau=0.1$, an object initially the size of Ceres will
grow at $1\,$AU (where $\xi \sim 1$), but not at $2\,$AU (where $\xi$
becomes large). This argument implies that, initially, all
planetesimals could have been the size of currently observed main belt
asteroids; those bodies in the asteroid belt did not grow appreciably,
while those at $1\,$AU did.  Therefore, we postulate that Mars's small
mass and the lack of planets in the asteroid belt might be the result
of this cutoff.  It is important to note that this figure just shows
$\xi$ and does not represent how the entire pebble accretion process
will behave.  In order to ascertain that, we must turn to numerical
calculations.  The remainder of this work presents such simulations.

\section{Methods}
To test this hypothesis, we performed a series of $N$-body
calculations of terrestrial planet formation starting with a
population of planetesimals embedded in our fiducial gas disk, adopted
from Ref$.$~\cite{Levison.etal.2015a}. We assume a flaring gas disk
with a surface density of $\Sigma = \Sigma_0 r_{\rm AU}^{-1}$
\cite{Andrews.etal.2010} and a scale height $h = 0.047~r_{\rm
  AU}^{9/7}~{\rm AU}$ \cite{Chiang.Goldreich.1997}, where $r_{\rm AU}$
is heliocentric distance in AU.  Here we set $\Sigma_0$ initially to
9000 g/cm$^2$, which is roughly 5 times that of the so-called minimum
mass solar nebula \cite{Hayashi.1981}.  In this work we set the
density of the planetesimals and pebbles, $\rho_s$, to 3 g cm$^{-3}$.
Additionally, our disk is assumed to be turbulent with $\alpha=3\times
10^{-4}$ \cite{Owen.etal.2011} (although see below).

We allowed the gas surface density to decrease exponentially with a
timescale of $t_g=2\,$Myr, which is motivated by observations
\cite{Haisch.etal.2001}.  The disk has solar composition so that the
solid-to-gas ratio is 0.005 in the terrestrial planet region
\cite{Lodders.2003}.  We convert a fraction $f_{pl}$ of the solids
into planetesimals at the beginning of the simulations.  We draw
  our planetesimals from a distribution of radii, $s$, of the form
$dN/ds \propto s^{-3.5}$ such that $s$ is between $s_l$ and $s_u$.
The values of $s_l$ and $s_u$ are assumed to be independent of
semi-major axis.  We set our fiducial value of $s_u$ to the
  radius of Ceres, $450\,$km, because we expect little growth in the
  asteroid belt and we need to produce this object.  However, we do
  vary $s_u$ from 100 to $600\,$km to test the sensitivity of our
  results to this value.  Because we are interested in building the
terrestrial planets and using the asteroid belt as a constraint, we
study the growth of planetesimals spread from 0.7 to $2.7\,$AU (the
presumed location of the snow line; \cite{Hayashi.1981}).  As is
typical for this type of simulation \cite{Raymond.etal.2009}, we do
not treat planet formation in the Mercury forming region in order to
save computer time.

\begin{figure}
	\centerline{\includegraphics[width=.5\textwidth]{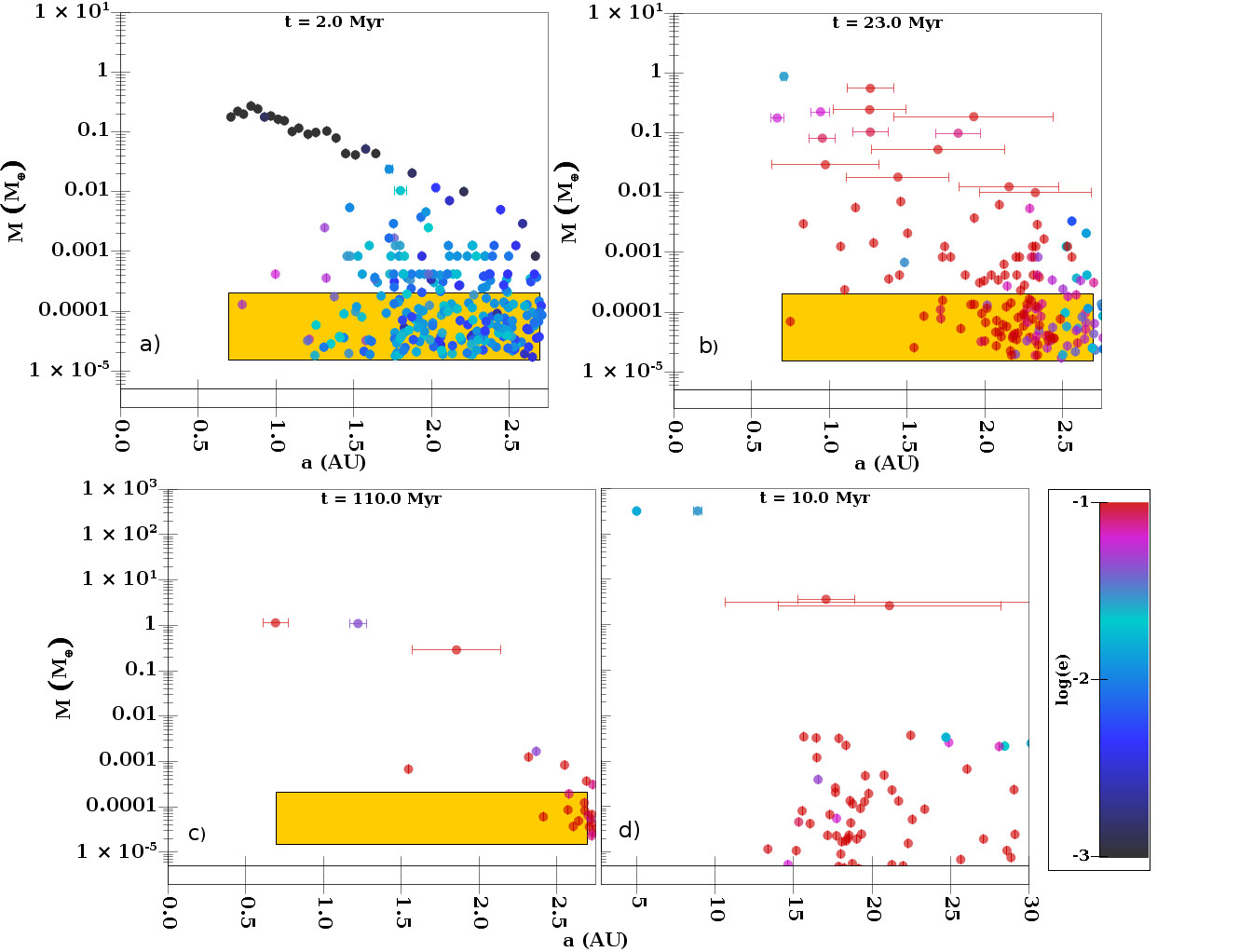}}
	\caption{Panels (a) to (c) show the formation of a system of terrestrial
		planets.  Each panel shows the mass of the planetesimals and planetary
		embryos as a function of semi-major axis, while color indicates their
		eccentricity.  In addition, the `error bars', which are only shown for
		objects larger than $0.01\,M_\oplus$ in order to decrease clutter,
		indicate the range of heliocentric distance that an object travels.
		The yellow box shows the region populated by planetesimals at $t=0$.
		Panel (b) shows the distribution of planetary embryos after pebble
		accretion but before the dynamical instability.  Panel (c) panel shows
		the final system.  The panel (d) shows the fiducial giant planet system
constructed in Ref.~\cite{Levison.etal.2015a}.}  \label{fig:ar}
\end{figure}

The remaining solids (assumed to be dust) are slowly converted
  into pebbles with a fixed initial $\tau$, which is a free parameter
  in our simulations.  Following Ref$.$~\cite{Levison.etal.2015a}, we
  utilize a simple prescription to convert dust into pebbles over time
  that assumes that pebbles form at a rate proportional 
   to the instantaneous dust mass, correcting for dust lost as the gas disk
	  evolves and as pebbles form. The functional form of pebble production can be found in Eq.~9 of the Methods section of Ref$.$~\cite{Levison.etal.2015a}.
 We scale the function such that the median production timescale is roughly
  $700,0000\,{\rm yr}$.  We assume that all of the pebbles are produced in 2
  Myr.  For simplicity, we assume that pebbles are randomly created throughout
  the disk according to the surface density.

We are justified in pursuing the long timescales for pebble
  formation for the following reasons.  While models of dust
  coagulation (Ref$.$~\cite{Brauer.etal.2008}, for example) predict
  that pebbles should grow on timescales on the order of 100-1000
  orbital periods, this result is observationally problematic.
  Millimeter and even centimeter sized particles, which should have
  been lost rapidly, are observed in disks of a range of ages (e.g$.$
  Ref.\cite{Ricci.etal.2010}).  While it is possible that the drift of
  these pebbles could be slowed by variations in the disk structure
  \cite{Johansen.etal.2009}, these trapping models need large, as of
  yet unobserved, variations in the disk structure.  An simpler
  alternative is that pebbles are continuously formed.  Indeed, models
  in which pebbles slowly form from dust and then are lost due to
  drift matches some features of observed disks
  \cite{Birnstiel.etal.2012}.  Therefore, we will assume an initial
  planetesimal population along with pebbles that are steadily
  produced by the disk over its lifetime.

We also assume that the pebbles involved in terrestrial planet
  formation formed within $2.7\,$AU.  By having a cutoff at this
  location, we are assuming that material drifting in from the outer
Solar System is unable to penetrate the snow-line, presumably due to
sublimation.  But no matter the mechanism, this assumption is
  required because solids from the outer Solar System are too carbon
  rich to have contributed more than a few percent of the mass of the
  terrestrial planets (see Ref.~\cite{Jessberger.etal.2001}).  Also,
  carbon can not be removed without heating the material to above
  $\sim$ 500K \cite{Drain.1979}, a temperature not reached in the
  midplane until well within 1 AU in reasonable disk models.

The values of $\tau$ present during VSPA are dependent on which
planetesimal formation model we assume and range from $\sim\!10^{-3}$
to $\sim\!10^{-1}$\cite{Cuzzi.etal.2001, Youdin.Goodman.2005,
  Cuzzi.etal.2008, Johansen.etal.2007}.  Ideally, here we would prefer
to study pebble sizes that cover the complete range required by both
models.  However, the CPU time required to perform our calculations
increases drastically as $\tau$ decreases because of two effects.
First, the timestep required by our code scales with the pebble's
aerodynamic stopping time.  Thus, a smaller $\tau$ requires a smaller
timestep.  In addition, pebbles with small $\tau$'s have slower radial
drift velocities than their larger siblings, and this spend more time
in the calculation.  As a result, at any time there are more objects
present in the simulation that the code needs to deal with.  This
significantly increases the required CPU time per timestep.
Therefore, to keep the calculations tractable, we require $\tau$ to be
larger than $\sim0.01$.  This issue will be addressed again in the
Discussion Section below.

Each system is evolved for $110\,$Myr using a Lagrangian code,
  {\tt LIPAD} \cite{Levison.etal.2012}.  {\tt LIPAD} is the first
  particle-based (i.e. Lagrangian) code that can follow the
  collisional/accretional/dynamical evolution of a large number of
  sub-kilometer objects through the entire growth process to become
  planets.  It is built on top of the symplectic $N$-body algorithm
  known as SyMBA \cite{Duncan.etal.1998}. In order to handle the very
  large number of sub-kilometer objects required by these simulations,
  we introduce the concept of a tracer particle. Each tracer
  represents a large number of small bodies with roughly the same
  orbit and size, and is characterized by three numbers: the physical
  radius, the bulk density, and the total mass of the disk particles
  represented by the tracer.  {\tt LIPAD} employs statistical
  algorithms for viscous stirring, dynamical friction, and collisional
  damping among the tracers. The tracers mainly dynamically interact
  with the larger planetary mass objects via the normal $N$-body
  routines, which naturally follow changes in the trajectory of
  tracers due to the gravitational effects of the planets and {\it
    vice versa}.  When a body is determined to have suffered an
  impact, it is assigned a new radius according to the probabilistic
  outcome of the collision based on a fragmentation law for basalt by
  Ref.~\cite{Benz.Asphaug.1999}.  In this way, the conglomeration of
  tracers and full $N$-body objects represent the size distribution of
  the evolving population.  {\tt LIPAD} is therefore unique in its
  ability to accurately handle the mixing and redistribution of
  material due to gravitational encounters, including
  planetesimal-driven migration and resonant trapping, while also
  following the fragmentation and growth of bodies.  An extensive
  suite of tests of {\tt LIPAD} can be found in
  Ref.~\cite{Levison.etal.2012}.  For the calculations described here,
  we will use a version of {\tt LIPAD} that has been modified to
  handle the particular needs of pebble accretion
  \cite{Kretke.Levison.2014}.

The calculations are performed in three stages.  For the first
$3\,$Myr, the terrestrial planet region is evolved in isolation as
pebbles continually form and drift inward.  At the end of this first
stage, all pebbles have either been accreted or lost, and thus no more
mass will be added to the system.  Because we are interested in
constructing systems similar to the Solar System, we only continue
simulations with the appropriate amount of material (between
$2.1\,M_\oplus$ and $2.7\,M_\oplus$) inside $2\,$AU.  If this
criterion is met, the simulation is cloned 6 times by adding a random
number between $-10^{-4}$ to $10^{-4}\,$AU to each component of the
position vector of each body.  For this second stage, we also add
Jupiter and Saturn in orbits consistent with their pre-migrated
configuration \cite{Morbidelli.etal.2007}.  In particular, they are
placed in their mutual 3:2 mean motion resonance with Jupiter at
$5.2\,$AU and Saturn at $8.6\,$AU.  The evolution is then followed
until $100\,$Myr.  For the final stage, Jupiter and Saturn are moved
to their current orbits and the system was integrated for an
additional $10\,$Myr.

\section{Simulation Results}
In total, we performed 28 simulations to at least $3\,$Myr, varying
$f_{pl}$ between 0.004 and 0.01, and $s_u$ between 100 to $600\,$km.
For comparison, note that asteroid (1) Ceres has a radius of
$476\,$km. Our small values for $f_{pl}$ were driven by the fact
  that we wanted to create asteroid belts as close to the observed
  system as possible assuming that VSPA is not effective there.
  Unfortunately, however, the smaller we force $f_{pl}$ the more
  tracers that we needed to resolve the system and thus the more CPU
  time required by the simulation.  Our compromise was to choose
  $f_{pl}$ so that the initial mass between 2.2 and $4\,$AU is roughly
  between 20 and 50 times the mass currently observed there. These
small values of $f_{pl}$ imply that there was at most $0.2\,M_\oplus$
of planetesimals between 0.7 and $2.7\,$AU; most of the mass in the
final planets comes from accreting pebbles.

We followed 9 of the 28 simulations to completion, making a total of
54 systems including the clones.  
See the {\it Supporting Information} for a description of the statistics of our runs.

The evolution of a system that produced a reasonable Solar System
analog is shown in Fig$.$~\ref{fig:ar}.  For this simulation, $\tau =
0.1$, $s$ is initially between between 200 and $450\,$km, and $f_{pl}
= 0.01$.  Growth occurs first and fastest closest to the Sun.  This
happens for two reasons.  As described above, $\sigma_{\rm peb}$ is
function of semi-major axis.  In addition, although we create pebbles
randomly throughout our computational domain following the surface
density of the gas disk, these objects quickly spiral toward the Sun
due to the effects of aerodynamic drag.  As a result, a planetesimal
encounters pebbles that were created outward of its semi-major axis
and thus the planetesimals closest to the Sun encounter more pebbles.

The era of pebble accretion ends when all the pebbles have been
generated and have either impacted the Sun or been accreted by an
embryo.  At this time (Fig$.$~\ref{fig:ar}a) there is a series of 23
embryos with masses greater than $0.01\,M_\oplus$ on quasi-stable,
nearly circular orbits.  There is a direct correlation between mass
and semi-major axis at this time, with no object larger than Mars
beyond $1.3\,$AU and none larger than the Moon beyond $1.9\,$AU.  The
largest object in the system is $0.27\,M_\oplus$.  This correlation
leads to very little mass beyond $1\,$AU.  As expected from
Fig$.$~\ref{fig:XI}, there was little growth beyond $2\,$AU.

This system remains stable until $20\,$Myr, at which time the orbits
of its embryos cross and they accrete each other. This dynamical
instability leads to the formation of two roughly Earth-mass objects
at 0.7 and $1.2\,$AU.  It is during this instability that a 2.7
Mars-mass object is gravitationally scattered to $1.9\,$AU by its
larger siblings and is stabilized by gravitational interactions with
smaller objects found there.  This leads to the system shown in the
Fig.~\ref{fig:ar}c that contains analogs of the Earth, Venus, and Mars
with roughly the correct masses and orbits.  
The basic evolution seen here, where the system first develops a
series of small planets on nearly circular orbits that suffer an
instability at a few tens of millions of years, is a common outcome in
our simulations.  Thus, this model predicts that the Solar System had
an initial generation of terrestrial planets --- consisting of a
large number of small planets --- that is now lost. Mars is likely a
remnant of this early system.  The late timing of the impact that
formed the moon \cite{Touboul.etal.2007} resulted, in part, from this
instability.

\begin{figure*}
	\includegraphics[width=\textwidth]{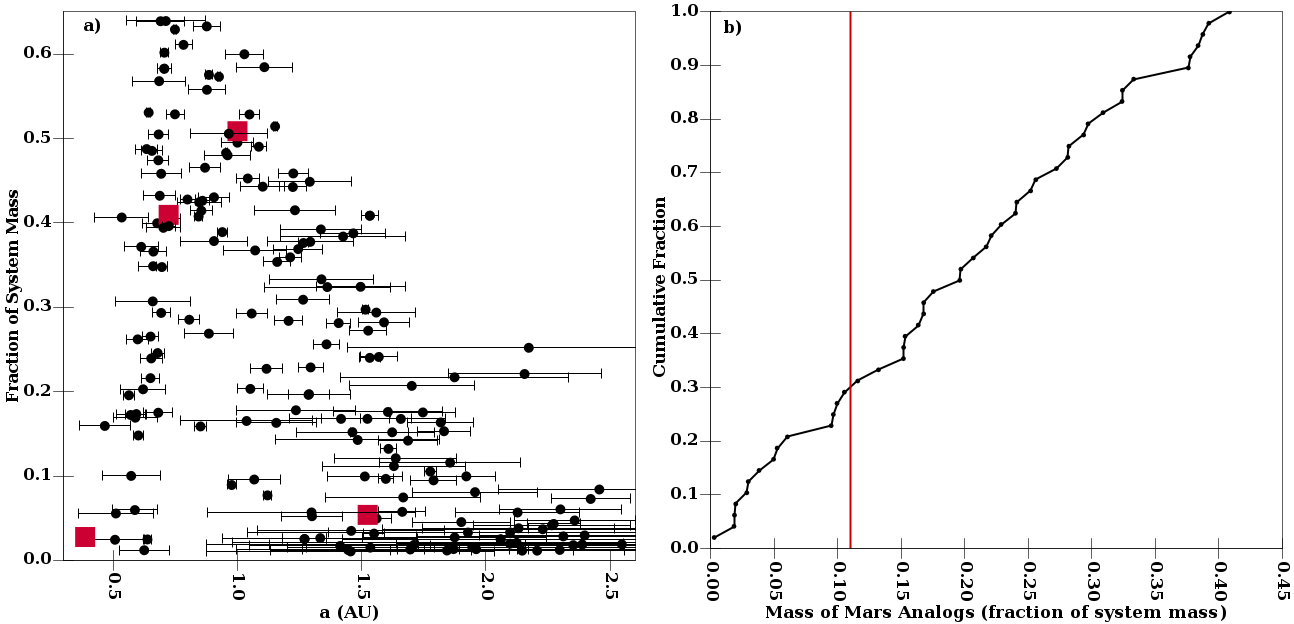}
    \caption{The finial distribution of our planets. a) The
      black dots show a compilation of the planets constructed during
      our simulations.  In particular, we plot the fraction of total
      system mass in each planet as a function of its semi-major axis.
      Here we plot the `fraction of total system mass' because
        the total mass of our systems varies from run to run due to
        the stochastic variations in the efficiency of pebble
        accretion. The `error bars' indicate the range of
      heliocentric distance that a planet travels and thus are a
      function of eccentricity.  The red squares indicate the real
      terrestrial planets in the Solar System. b) The cumulative
        mass distribution of Mars analogs.  Planets to the left of the
        red line are considered good Mars analogs according to
        Ref.~\cite{Fischer.Ciesla.2014}.}  \label{fig:comp}
\end{figure*}

Although, due to the chaotic nature of planet formation, not all of
our simulations produce good Solar System analogs, as shown in
Fig.~\ref{fig:comp}a.  While interpreting this plot it must be
noted that we did not make any attempt to uniformly cover parameter
space.  Each calculation took many weeks to perform and thus the
survey of parameter space was limited and {\it ad hoc}.  Thus, the
observed distribution shown needs to be viewed with caution.  However,
we believe that some of the trends are robust.  For example, it is
common to produce reasonable Earth and Venus analogs.  In addition,
planets near $1.5\,$AU are systematically smaller than their siblings
interior to $1\,$AU.  No planets grow beyond $2\,$AU.  The objects
seen in this region were scattered out during a dynamical instability.
The majority of these objects are still on orbits that cross their
larger neighbors and thus will eventually be removed.  However, a few
are on stable orbits.  The anomalous and highly processed main belt
asteroid (4) Vesta might be an object that was captured in this
manner.

The natural question is whether VSPA produces good Mars analogs
  more frequently than the standard planet formation picture.
  Ref.~\cite{Fischer.Ciesla.2014} finds that the standard model
  produces a good Mars analog in 4\% of the simulations with Jupiter
  and Saturn on circular orbits (their likely state at the time of
  terrestrial planet formation).  They define a `Mars analog' as the
  largest planet with a semi-major axis both in the range 1.25 --
  $2\,$AU and outside of the Earth analog's orbit. They define an
  `Earth analog' to be the largest planet between 0.75 and $1.25\,$AU
  and if there is no planet within this range, the Earth analog is the
  closest planet to $1\,$AU. A `good' Mars analog is defined to be an
  object that is smaller than $0.22\,M_\oplus$ or 11\% of the total
  system mass. Keeping the above caveats about our parameter space
  coverage in mind, we plot the cumulative mass distribution of Mars
  analogs from our simulations in Fig.~\ref{fig:comp}b.  We find that
  30\% of our systems produce good analogs.  This represents a
  significant improvement over the standard model. Moreover,
  Ref.~\cite{Fischer.Ciesla.2014} finds that the probability that the
  standard model produces systems with both a good Mars analog and a
  low mass asteroid belt is roughly 0.6\%.  By design, our systems
  always have a low mass asteroid belt.

\section{Discussion}
The above experiments show that Mars's small mass is a natural
  outcome of the process of VSPA.  Having said this, there are other
  issues that need to be considered.

\subsection{Sizes of Pebbles} There are currently two
  independent models in the literature for planetesimal formation
  directly from pebbles that make very different predictions
  concerning the size of pebbles: the streaming instability and
  turbulent concentration.  The streaming instability takes advantage
  of the fact that a clump of pebbles is less affected by aerodynamic
  drag, and thus moves at a different speed, than free-floating
  pebbles \cite{Youdin.Goodman.2005}.  As a result, individual pebbles
  catch up with the clump, and the clump grows to the point where it
  becomes gravitationally bound and collapse to form a planetesimal
  \cite{Johansen.etal.2007}.  This mechanism requires $\tau \sim 0.1$
  -- $1$ \cite{Bai.Stone.2010}.  On the other hand, turbulent
  concentration scenarios suggest that objects with $\tau \sim
  10^{-3}$ migrate to the outer edges of turbulent eddies due to
  centrifugal forces caused by the eddie's rotation
  \cite{Cuzzi.etal.2001,Cuzzi.etal.2008}.  There the pebbles form
  gravitationally bound clumps that collapse to create planetesimals.
  While we are agnostic on which of these are correct and would prefer
  to cover the whole range of values, as we described above, we cannot
  perform simulations with $\tau$'s smaller than $\sim\!10^{-2}$
  because the required amount of CPU time makes these calculations
  impractical.

Unfortunately, there have been recent developments in our
understanding of the coagulation of pebbles that calls into question
whether objects with $\tau \sim 0.01$ can grow in disks.  In
traditional models the growth of objects is limited by particle
fragmentation \cite{Birnstiel.etal.2012}.  With this limitation,
particles should grow to $\tau \sim 0.1$ given the known strengths of
compacted silicate grains \cite{Wurm.etal.2005}.  However, some
coagulation simulations have identified another barrier to growth,
known as the ``bouncing barrier,'' that halts growth of rocky pebbles
at much small sizes \cite{Zsom.etal.2010,Guttler.etal.2010}.  If this
barrier is proven robust then pebble growth may halt at $\tau \sim
10^{-4}-10^{-3}$.

The fact that we cannot perform direct {\tt LIPAD} calculations with
$\tau \sim 10^{-3}$ does not imply that our basic model will not work
in this regime.  For example, the simple calculation shown in
Fig$.$~\ref{fig:XI} shows that, regardless of pebble size, there is a
radial dependence on the efficiency of pebble accretion.  The only
difference is that the transition between efficient and inefficient
growth occurs at much smaller planetesimal size.  This suggests that
we would be able to create the same general features seen in
Fig$.$~\ref{fig:ar}c (a low mass Mars and asteroid belt) for a range
of assumed pebble sizes as long as a decrease in pebble size is
commensurate with a decrease in the maximum initial planetesimal size.

This trend is already apparent with the range of pebble size we can
investigate with {\tt LIPAD} (see the table in the {\it Supporting
  Information}).  For example, when $\tau$ was $0.08$, reasonable
terrestrial analogs were found only when the largest planetesimals
initially in the system were bigger than $\sim\!300\,$km.  However
when $\tau=0.025$, simulations that initially contained planetesimals
this big grew planets in the asteroid belt.  For $\tau$'s of this
value, systems that look like our Solar System occurred only when that
the maximum initial planetesimal size was $\sim\!100\,$km. Indeed,
recent work indicates that the size-distribution of the larger
asteroids in the main belt could be explained by planetesimals of
roughly this size accreting $\tau \sim 10^{-3}$
pebbles\cite{Johansen.etal.2015}.  Therefore, we expect that an
initial combination of $\tau \sim 10^{-3}$ pebbles and planetesimals
of a few tens of kilometers would result in systems like our own.

Having said this, even if the bouncing barrier model is correct, it
has some uncertainties and free parameters that might allow larger
pebbles to grow.  One of the most important being the assumed level of
turbulence in the disk.  In environments with very little turbulence
($\alpha \sim 10^{-6}-10^{-5}$) it has been demonstrated that small
pebbles and dust can combine to form larger aggregates with $\tau\sim
10^{-1}$ \cite{Ormel.etal.2008}.
  
In our simulations we chose $\alpha = 3\times 10^{-4}$, a value motivated by
disk evolution timescales \cite{Haisch.etal.2001} with the assumption that
uniform turbulence is driving angular momentum transport in protoplanetary
disks.  However, while angular momentum must be being transferred in these
accretion disks, this does not necessarily imply that there is turbulence at
the midplane where the pebbles and planetesimals are located.  Indeed, recent
models of angular momentum transport in disks suggest that turbulence may be
limited to the upper and lower regions of the disk \cite{Gammie.1996} or that
disks may not be turbulent at all, and angular momentum may instead by
transferred by magnetocentrifulgal disk winds \cite{Bai.2014}.  Therefore we
investigated how pebble accretion would behave in more quiescent disks that may
be more consistent with our assumed pebble size.  We find that the masses
of the growing planetary embryos are independent of the assumed level of
turbulence for $\alpha$ ranging from $10^{-6}$ to our fiducial value of
$3\times 10^{-4}$ (see Fig.~4).
As a result, although we did not directly perform
simulations in a low-$\alpha$ environment where $\tau\sim 10^{-1}$
pebbles can form even if the bouncing barrier is important, our
results are directly applicable there.

\begin{figure}
	\includegraphics[width=0.5\textwidth]{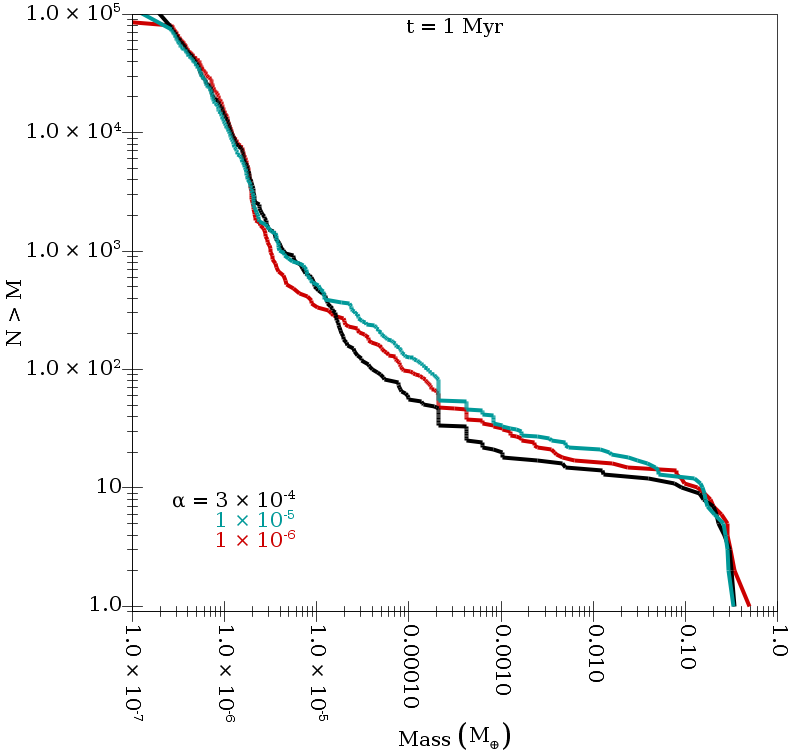}
\caption{\label{fig:alpha} The cumulative mass distribution of objects
  at the end of pebble accretion (i.e$.$ at $1\,$Myr) in three
  simulations that are identical except for $\alpha$.  These runs have
  $f_{pl} = 0.004$, $s = 50$ -- $100\,$km, and $\tau= 0.025$ (our
  smallest value).  Three values of $\alpha$ where studied; $\alpha =
  3 \times 10^{-4}$ (our fiducial value; black curve), $10^{-5}$ (cyan
  curve), and $10^{-5}$ (red curve).  Note that all three curves are
  basically identical.}
\end{figure}

\subsection{Implications for the Asteroid Belt}
This model also has profound implications for both the location
  and history of the asteroid belt.  One of the primary reasons
  why $\xi$ increases with heliocentric distance in
  Fig$.$~\ref{fig:XI} is because we assume the disk flares (i.e$.$ the
  ratio between the gas disk scale height and the heliocentric
  distance increases with heliocentric distance). While it is observed
  that most disks are flaring in the outer regions
  \cite{Kenyon.Hartmann.1987}, there are not direct observations
  constraining the disk shape in the terrestrial planet region.
  Indeed, if a disk is viscously heated and one were to assume that
  the opacity of the disk is constant with heliocentric distance, then
  the disk will be flat, with a constant aspect ratio (e.g$.$
  \cite{Garaud.Lin.2007}). However, full radiative transfer models of
  the structure of viscously heated accretion disks (e.g$.$
  \cite{Bitsch.etal.2014}) show flaring interior to the snow line
  (which is why we made the assumption we did).  This occurs because
  the dust opacity is expected to increase with decreasing
  temperatures as water freezes out.  In viscously heated disk this
  occurs interior to the mid-plane snow-line because the temperture
  decreases with height.  The change in opacity causes the disk to
  flare.  Thus, a region of increasing $\xi$ is likely associated with
  snow-lines, and VSPA might lead us to expect that asteroid belts are
  features that are naturally associated with regions interior to
  snow-lines.

Our VSPA calculations also produce asteroid belts are profoundly
  different than the standard model. In the standard model, the
region between 2 and $2.7\,$AU originally contained several
Earth-masses of planetesimals --- $\gtrsim\!99.9\%$ of which were lost
either through the dynamical effects of Jupiter and Saturn or by
collisional grinding \cite{Lecar.Franklin.1973}(the asteroid belt
currently contains $\sim\!5\times 10^{-4}\,M_\oplus$).  Since $f_{pl}$
is always less than 0.1 in our simulations and little growth occurred
beyond $\sim\!2\,$AU, there was never very much mass in the asteroid
belt.  Indeed, our final planetary systems have asteroid belts with
masses between 0.5 and 155 of that currently observed in this region.
The model in Fig.~\ref{fig:ar} has an asteroid belt 3.5 times more
massive than the current belt.  This is a reasonable value because we
expect the asteroid belt to lose between roughly $50\%$ and $90\%$ of
its mass due to chaotic diffusion, giant planet migration, and
collisions over the age of the Solar System
\cite{Walsh.etal.2011,Walsh.etal.2012}.

As previously noted by Ref.~\cite{Walsh.etal.2011}, an early low-mass
asteroid belt might explain several of its most puzzling
characteristics.  For example, a low-mass main belt is consistent with
a limited degree of collisional evolution, as predicted by modeling
work \cite{Bottke.etal.2005}.  Similarly, the 530 km diameter
differentiated asteroid (4) Vesta only has two 400--$500\,$km diameter
basins on its surface, and one of them is only $\sim$1 Gy old
\cite{Marchi.etal.2012}.  These data are a good match to a primordial
main belt population that was never very massive.  Finally, if
  Vesta formed as part of a primordial asteroid belt that contained
  hundreds of times the mass of the current population, then it would
  be statistically likely have hundreds of Vesta-like bodies as
well. Even a short interval of collisional evolution would produce
more basaltic fragments than can be accommodated by the existing
collisionally and dynamically evolved main belt \cite{Bottke.2014}.
All of these observables could be explained by asteroid belts that
never contained much mass, like those constructed in our models.

Additionally, we note that by varying the parameters in this model,
particularly the initial planetesimal size and the pebble Stokes
number, we produce a wide range of different planetary architectures.
While we present parameters that produce systems similar to our own,
the natural variability of this process leads us to speculate that
VSPA might be able to explain the variety in the observed exoplanetary
systems \cite{Batalha.etal.2013}.

\subsection{Unified Picture of the Solar System Formation}
Finally for completeness, Fig$.$~\ref{fig:ar}d shows a giant planet
system constructed within the same disk and using VSPA
\cite{Levison.etal.2015a}. It shows two gas giant planets plus 3 ice
giants. There is no growth beyond $20\,$AU because $\xi$ is again
large.

We note that we get growth in the giant planet region
\cite{Levison.etal.2015a} and not in the asteroid belt for at least
three reasons: 1) The initial planetesimals were probably larger in
the outer Solar System (to be consistent with the larger sizes of
object in the Kuiper Belt as compared to the asteroid belt).  In
particular, in this work we use a maximum planetesimal size in the
terrestrial region based on Ceres, while in
Ref.~\cite{Levison.etal.2015a} we used a maximum planetesimal size
based on Pluto in the giant planet region.  2) The pebble sizes were
likely also bigger because because the pebbles are icy and thus
stickier \cite{Gundlach.Blum.2015}.  3) the giant planets have access
to a much larger reservoir of pebbles than the inner Solar System due
to sublimation of ices at the snow line. In particular as we
  explained above, here we assume that pebbles from the outer Solar
  System do not contribute to the growth of the terrestrial planets.
  As a result, the terrestrial planets only have access to solids out
  to $2.7\,$AU.  In Ref.~\cite{Levison.etal.2015a}, we assumed that
  pebbles formed out to $30\,$AU and thus the giant planets had access
  to the substantially larger amount of solids that were between 2.7
  and $30\,$AU.

The lower two panels in the figure, therefore show a consistent
planetary system generated by VSPA.  This model reproduces the basic
structure of our entire planetary system --- two roughly Earth mass
objects between 0.5 and $1\,$AU, a small Mars, a low mass asteroid
belt, two gas giants, ice giants, and a primordial Kuiper belt that
contains objects the mass of (134340) Pluto and (136199) Eris but did
not form planets.

\begin{acknowledgments}
  This research was supported by the NASA
  Solar System Exploration Research Virtual Institute.  We would like
  to thank M.~Duncan, S.~Jacobson, M.~Lambrechts, A.~Morbidelli, and
  D.~Nesvorn{\' y}
  for useful discussions.
\end{acknowledgments}

\section{Statistics of Our Simulations}

Tables~S\ref{tab:runs} \& S\ref{tab:runsb} list the simulations that we
have completed in our investigation of terrestrial planet formation.
All of these employed the following setup.  We used a total surface
density distribution of $\Sigma = \Sigma_0 r_{\rm AU}^{-1}$, where
$r_{\rm AU}$ is the heliocentric distance in AU, and $\Sigma_0 =
9000\, e^{-t/2{\rm Myr}}~~{\rm g/cm}^2$.  The disk flares with a scale
height of $h = 0.047~r_{\rm AU}^{9/7}~{\rm AU}$
\cite{Chiang.Goldreich.1997}.  We assume a solid-to-gas ratio of
0.005 and a bulk density of 3 g/cm$^3$.  Pebbles are slowly
generated and follow the same spatial distribution as $\Sigma$ above.
In particular, since pebble generation follows the evolution of the
gas disk, the median time for which a pebble is generated is
$\sim\!0.7\,$Myr.

There are three important free parameters that we varied.  These are
listed in the first three columns of the tables:

\begin{enumerate}

\item{} The fraction of solids in the disk that are initially
  converted into planetesimals, $f_{pl}$.  For the fiducial simulation, $f_{pl}= 0.01$.

\item{} The range of initial radii of the planetesimals. We draw the
  initial planetesimals from a distribution of radii, $s$, of the form
  $dN/ds \propto s^{-3.5}$ \cite{Morbidelli.etal.2009}. For our fiducial
  simulation $200 \leq s \leq 450\,$km.

\item{} The initial Stokes number of a pebble, $\tau$.  Note that as pebbles
	spiral toward the Sun, their Stokes number changes (generally
	decreases) because both the gas density and the orbital period vary.  For
	our fiducial simulation $\tau = 0.1$.

\end{enumerate}

\noindent The last four columns in the table present the basic
characteristics of our systems.  In particular, $M(<2\,{\rm AU})$ is
the total mass found within $2\,$AU at $3\,$Myr.  Recall that we only
continued to integrate systems for which $2.1 \leq M(<2\,{\rm AU})
\leq 2.8\,M_\oplus$.  We chose this range because it is slightly larger
the total of the terrestrial planets ($2.0\,M_\oplus$) and thus
allow for some subsequent loss.  We created 6 clones of any run that satisfies
the above criterion.  The clone ID is listed in Column~(5).
Column~(6) lists the total mass in the asteroid belt (here defined to
be between 2.1 and $5\,$AU) in terms of the currently observed
asteroid belt mass ($M_{\rm AB}$).  Finally Columns~(7) and (8) list
the total number of planets with mass between 0.5 Mars-mass and
$0.5\,M_\oplus$ ($N_{\rm Mars}$) and the number with mass greater than
$0.5\,M_\oplus$ ($N_{\rm Earth}$), respectively.

%\begin{thebibliography}{1}
%\setcounter{enumiv}{16}
%
%\bibitem{Chiang.Goldreich.1997}
%Chiang EI, Goldreich P (1997) {Spectral Energy Distributions of T Tauri Stars
%  with Passive Circumstellar Disks}.
%\newblock {\em \apj} 490:368--376.
%\setcounter{enumiv}{52}
%
%\bibitem{Morbidelli.etal.2009}
%Morbidelli A, Bottke WF, Nesvorn\'{y} D, Levison HF (2009) {Asteroids were born
%  big}.
%\newblock {\em \icarus} 204:558--573.

%\end{thebibliography}
\setcounter{table}{0}

\begin{table}
	%\caption{AHHHH!} \label{ahhh}
\caption{Our completed simulations.} \label{tab:runs}
\end{table}

\setcounter{table}{1}
\begin{table*}[ht]
\centering
%\caption{Our completed simulations.} \label{tab:runs}
\caption{Our completed simulations (cont).} \label{tab:runsb}
\begin{tabular}{@{\extracolsep{\fill}}cccccccc}
\hline
& &  &  &  &  &  & \\
    (1)     & (2)&  (3)    & (4)   & (5)   &  (6)    & (7) & (8) \\     
$f_{pl}$ & $s$ & $\tau$ &  $M(<2\,{\rm AU})$ & & $M (>2.1\,{\rm AU})$ & $N_{\rm Mars}$ & $N_{\rm Earth}$ \\

 &  (km) &  & ($M_\oplus$)  &  &  ($M_{\rm AB}$) & & \\
 & &  &  &  &  &  & \\
\hline
 & &  &  &  &  &  & \\
0.004 & 50-100km & 0.02 &  3.3 &  &  &  &  \\
0.004 & 50-100km & 0.025 &  2.4 &  &  &  &  \\
 &  &  & & Clone 1  & 13 & 1 & 2 \\
 &  &  & & Clone 2  & 13 & 0 & 2 \\
 &  &  & & Clone 3  & 12 & 0 & 2 \\
 &  &  & & Clone 4  & 37 & 1 & 1 \\
 &  &  & & Clone 5  & 7 & 0 & 2 \\
 &  &  & & Clone 6  & 11 & 0 & 2 \\
0.004 & 50-100km & 0.03 &  1.6 &  &  &  &  \\
0.004 & 50-100km & 0.04 &  1.3 &  &  &  &  \\
0.004 & 50-100km & 0.06 &  1.1 &  &  &  &  \\
0.004 & 100-300km & 0.03 &  3.6 &  &  &  &  \\
0.004 & 100-300km & 0.06 &  3.1 &  &  &  &  \\
0.004 & 100-300km & 0.08 &  2.4 &  &  &  &  \\
 &  &  & & Clone 1  & 1.2 & 2 & 1 \\
 &  &  & & Clone 2  & 11 & 1 & 3 \\
 &  &  & & Clone 3  & 82 & 2 & 2 \\
 &  &  & & Clone 4  & 112 & 1 & 4 \\
 &  &  & & Clone 5  & 31 & 1 & 3 \\
 &  &  & & Clone 6  & 37 & 0 & 2 \\
0.004 & 100-300km & 0.1 &   1.7 & &  &  &  \\
0.004 & 200-450km & 0.03 &  3.9 & &  &  &  \\
0.004 & 200-450km & 0.06 &  3.5 & &  &  &  \\
0.004 & 200-450km & 0.08 &  2.8 & &  &  &  \\
 &  &  & & Clone 1  & 2 & 0 & 2 \\
 &  &  & & Clone 2  & 7 & 2 & 2 \\
 &  &  & & Clone 3  & 5 & 1 & 2 \\
 &  &  & & Clone 4  & 10 & 0 & 3 \\
 &  &  & & Clone 5  & 17 & 0 & 3 \\
 &  &  & & Clone 6  & 50 & 2 & 2 \\
0.004 & 200-450km & 0.1 &  2.1 & &  &  &  \\
 &  &  & & Clone 1  & 81 & 0 & 2 \\
 &  &  & & Clone 2  & 12 & 1 & 3 \\
 &  &  & & Clone 3  & 72 & 2 & 2 \\
 &  &  & & Clone 4  & 9 & 3 & 1 \\
 &  &  & & Clone 5  & 18 & 0 & 2 \\
 &  &  & & Clone 6  & 6 & 2 & 2 \\
 & &  &  &  &  &  & \\
\hline
\end{tabular}
\end{table*}

\setcounter{table}{2}
\begin{table*}[ht]
\centering
\caption{Our completed simulations (cont).} \label{tab:runsb}
\begin{tabular}{@{\extracolsep{\fill}}cccccccc}
\hline
 & &  &  &  &  &  & \\
    (1)     & (2)&  (3)    & (4)   & (5)   &  (6)    & (7) & (8) \\     

$f_{pl}$ & $s$ & $\tau$ &  $M(<2\,{\rm AU})$ & & $M (>2.1\,{\rm AU})$ & $N_{\rm Mars}$ & $N_{\rm Earth}$ \\

 &  (km) &  & ($M_\oplus$)  &  &  ($M_{\rm AB}$) & & \\
 & &  &  &  &  &  & \\
\hline
 & &  &  &  &  &  & \\
0.008 & 100-300km & 0.03 &  5.7 & &  &  &  \\
0.008 & 100-300km & 0.06 &  3.1 & &   &  &  \\
0.008 & 100-300km & 0.08 &  2.4 & &  &  &  \\
 &  &  & & Clone 1  & 82 & 3 & 1 \\
 &  &  & & Clone 2  & 32 & 2 & 2 \\
 &  &  & & Clone 3  & 110 & 4 & 1 \\
 &  &  & & Clone 4 & 31 & 2 & 2 \\
 &  &  & & Clone 5  & 17 & 3 & 1 \\
 &  &  & & Clone 6  & 19 & 3 & 1 \\
0.008 & 100-300km & 0.1 &     2 &  & &  &  \\
0.008 & 200-450km & 0.03 &  5.9 & &   &  &  \\
0.008 & 200-450km & 0.06 &  3.8 & &  &  &  \\
0.008 & 200-450km & 0.1 &   2.7 & &  &  &  \\
 &  &  & & Clone 1  & 182 & 3 & 2 \\
 &  &  & & Clone 2  & 12 & 2 & 2 \\
 &  &  & & Clone 3  & 15 &  &  \\
 &  &  & & Clone 4  & 15 & 1 & 2 \\
 &  &  & & Clone 5  & 224 & 2 & 2 \\
 &  &  & & Clone 6  & 113 & 3 & 2 \\
0.008 & 200-450km & 0.3 &  1.1 & &  &  &  \\
0.008 & 200-600km & 0.1 &  2.6 & &  &  &  \\
 &  &  & & Clone 1  & 57 & 1 & 2 \\
 &  &  & & Clone 2  & 208 & 3 & 2 \\
 &  &  & & Clone 3  & 16 & 2 & 1 \\
 &  &  & & Clone 4  & 42 & 2 & 1 \\
 &  &  & & Clone 5  & 22 & 3 & 0 \\
 &  &  & & Clone 6  & 17 & 3 & 2 \\
0.01 & 100-300km & 0.06 &  3.6 & &   &  &  \\
0.01 & 100-300km & 0.08 &  3.1 & &  &  &  \\
0.01 & 100-300km & 0.1 &  2.6 & &  &  &  \\
 &  &  & & Clone 1  & 0.5 & 0 & 2 \\
 &  &  & & Clone 2  & 12 & 1 & 2 \\
 &  &  & & Clone 3  & 102 & 1 & 2 \\
 &  &  & & Clone 4  & 155 & 2 & 3 \\
 &  &  & & Clone 5  & 19 & 0 & 2 \\
 &  &  & & Clone 6  & 28 & 1 & 3 \\
0.01 & 200-450km & 0.03   & 6.3 & &  &  &  \\
0.01 & 200-450km & 0.1   & 2.7 & &  &  &  \\
 &  &  & & Clone 1  & 7 & 2 & 2 \\
 &  &  & & Clone 2  & 52 & 2 & 3 \\
 &  &  & & Clone 3  & 54 & 0 & 2 \\
 &  &  & & Clone 4  & 19 & 1 & 2 \\
 &  &  & & Clone 5  & 38 & 1 & 2 \\
 &  &  & & Clone 6  & 16 & 1 & 2 \\
0.01 & 200-450km & 0.3 & 1.2 &  & &  &  \\
 & &  &  &  &  &  & \\
\hline
\end{tabular}
\end{table*}

\end{article}


\begin{thebibliography}{10}

\bibitem{Raymond.etal.2009}
{Raymond} SN, {O'Brien} DP, {Morbidelli} A, {Kaib} NA (2009) {Building the
  terrestrial planets: Constrained accretion in the inner Solar System}.
\newblock {\em \icarus} 203:644--662.

\bibitem{Hansen.2009} {Hansen}, BMS (2009) {Formation of the
  Terrestrial Planets from a Narrow Annulus}.
\newblock {\em \apj} 703:1131--1140.

\bibitem{Izidoro.etal.2014} {Izidoro} A, {Haghighipour} N,
  {Winter} OC, {Tsuchida} M (2014) {Terrestrial Planet Formation in
  a Protoplanetary Disk with a Local Mass Depletion: A Successful
  Scenario for the Formation of Mars}.
\newblock {\em \apj} 782:31.

\bibitem{Walsh.etal.2011}
{Walsh} KJ, {Morbidelli} A, {Raymond} SN, {O'Brien} DP, {Mandell} AM (2011) {A
  low mass for Mars from Jupiter's early gas-driven migration}.
\newblock {\em \nat} 475:206--209.

\bibitem{Lambrechts.Johansen.2012}
{Lambrechts} M, {Johansen} A (2012) {Rapid growth of gas-giant cores by pebble
  accretion}.
\newblock {\em \aap} 544:A32.

\bibitem{Levison.etal.2015a}
{Levison} HF, {Kretke} KA, {Duncan} M (2015) 
Growing the gas-giant planets by the gradual accumulation of pebbles.
\newblock {\em Nature} 524:32--324.

\bibitem{Cuzzi.etal.2001}
Cuzzi JN, Hogan RC, Paque JM, Dobrovolskis AR (2001) {Size-selective
  Concentration of Chondrules and Other Small Particles in Protoplanetary
  Nebula Turbulence}.
\newblock {\em \apj} 546:496--508.

\bibitem{Youdin.Goodman.2005}
Youdin AN, Goodman J (2005) {Streaming Instabilities in Protoplanetary Disks}.
\newblock {\em \apj} 620:459--469.

\bibitem{Cuzzi.etal.2008}
Cuzzi JN, Hogan RC, Shariff K (2008) {Toward Planetesimals: Dense Chondrule
  Clumps in the Protoplanetary Nebula}.
\newblock {\em \apj} 687:1432--1447.

\bibitem{Johansen.etal.2007}
Johansen A et~al. (2007) {Rapid planetesimal formation in turbulent
  circumstellar disks}.
\newblock {\em \nat} 448:1022--1025.

\bibitem{Goldreich.Ward.1973}
Goldreich P, Ward WR (1973) {The Formation of Planetesimals}.
\newblock {\em \apj} 183:1051--1062.

\bibitem{Youdin.2011a}
{Youdin} AN (2011) {On the Formation of Planetesimals Via Secular Gravitational
  Instabilities with Turbulent Stirring}.
\newblock {\em \apj} 731:99.

\bibitem{Ormel.Klahr.2010}
{Ormel} CW, {Klahr} HH (2010) {The effect of gas drag on the growth of
  protoplanets. Analytical expressions for the accretion of small bodies in
  laminar disks}.
\newblock {\em \aap} 520:A43.

\bibitem{Morbidelli.Nesvorny.2012}
{Morbidelli} A, {Nesvorny} D (2012) {Dynamics of pebbles in the vicinity of a
  growing planetary embryo: hydro-dynamical simulations}.
\newblock {\em \aap} 546:A18.

\bibitem{Bai.Stone.2010}
{Bai} XN, {Stone} JM (2010) {Dynamics of Solids in the Midplane of
  Protoplanetary Disks: Implications for Planetesimal Formation}.
\newblock {\em \apj} 722:1437--1459.

\bibitem{Andrews.etal.2010}
Andrews SM, Wilner DJ, Hughes AM, Qi C, Dullemond CP (2010) {Protoplanetary
  Disk Structures in Ophiuchus. II. Extension to Fainter Sources}.
\newblock {\em \apj} 723:1241--1254.

\bibitem{Chiang.Goldreich.1997}
Chiang EI, Goldreich P (1997) {Spectral Energy Distributions of T Tauri Stars
  with Passive Circumstellar Disks}.
\newblock {\em \apj} 490:368--376.

\bibitem{Hayashi.1981}
Hayashi C (1981) {Structure of the Solar Nebula, Growth and Decay of Magnetic
  Fields and Effects of Magnetic and Turbulent Viscosities on the Nebula}.
\newblock {\em Progress of Theoretical Physics Supplement} 70:35--53.

\bibitem{Owen.etal.2011}
{Owen} JE, {Ercolano} B, {Clarke} CJ (2011) {Protoplanetary disc evolution and
  dispersal: the implications of X-ray photoevaporation}.
\newblock {\em \mnras} 412:13--25.

\bibitem{Haisch.etal.2001}
{Haisch} KE, Lada EA, Lada CJ (2001) {Disk Frequencies and Lifetimes in Young
  Clusters}.
\newblock {\em \apjl} 553:L153--L156.

\bibitem{Lodders.2003}
Lodders K (2003) {Solar System Abundances and Condensation Temperatures of the
  Elements}.
\newblock {\em \apj} 591:1220--1247.

\bibitem{Brauer.etal.2008} Brauer F, Dullemond CP, Henning T
  (2008) Coagulation, fragmentation and radial motion of solid
  particles in protoplanetary disks.  
\newblock {\em \aap} 480:  859--877.

\bibitem{Ricci.etal.2010} Ricci L, Testi L, Natta A, Neri R, Cabrit S,
  Herczeg GJ (2010) {Dust properties of protoplanetary disks in the
    Taurus-Auriga star forming region from millimeter wavelengths}.
\newblock {\em \aap} 512:A15.

\bibitem{Johansen.etal.2009} {Johansen A}, {Youdin A}, {Mac Low}
  MM (2009) {Particle Clumping and Planetesimal Formation Depend Strongly on
    Metallicity}.  
\newblock \emph{\apjl} 704:L75--L79.

\bibitem{Birnstiel.etal.2009} {Birnstiel} T, Klahr H, Ercolano B
  (2009) {A simple model for the evolution of the dust population in
  protoplanetary disks}.  
\newblock \emph{\aap} 539:A148.

\bibitem{Jessberger.etal.2001}
{Jessberger} EK et~al. (2001) {\em {Properties of Interplanetary Dust:
  Information from Collected Samples}}, eds.{} {Gr{\"u}n} E, {Gustafson} BAS,
  {Dermott} S, {Fechtig} H.
\newblock p. 253.

\bibitem{Drain.1979} {Draine} BT (1979) On the chemisputtering of
  interstellar graphite grains.  \newblock {\em \apj} 230:106-115.

\bibitem{Levison.etal.2012}
{Levison} HF, {Duncan} MJ, {Thommes} E (2012) {A Lagrangian Integrator for
  Planetary Accretion and Dynamics (LIPAD)}.
\newblock {\em \aj} 144:119--138.

\bibitem{Duncan.etal.1998} Duncan MJ, Levison HF, Lee MH (1998) A
  Multiple Time Step Symplectic Algorithm for Integrating Close
  Encounters.
\newblock {\em \aj} 116:2067--2077.

\bibitem{Benz.Asphaug.1999} Benz W, Asphaug 
E  (1999) Catastrophic Disruptions Revisited.
\newblock {\em \icarus} 142:5--20. 


\bibitem{Kretke.Levison.2014}
{Kretke} KA, {Levison} HF (2014) {Challenges in Forming the Solar System's
  Giant Planet Cores via Pebble Accretion}.
\newblock {\em \aj} 148:109.

\bibitem{Morbidelli.etal.2007}
{Morbidelli} A, {Tsiganis} K, {Crida} A, {Levison} HF, {Gomes} R (2007)
  {Dynamics of the Giant Planets of the Solar System in the Gaseous
  Protoplanetary Disk and Their Relationship to the Current Orbital
  Architecture}.
\newblock {\em \aj} 134:1790--1798.

\bibitem{Birnstiel.etal.2012}
{Birnstiel} T, {Klahr} H, {Ercolano} B (2012) {A simple model for the evolution
  of the dust population in protoplanetary disks}.
\newblock {\em \aap} 539:A148.

\bibitem{Wurm.etal.2005}
{Wurm} G, {Paraskov} G, {Krauss} O (2005) {Growth of planetesimals by impacts
  at $\sim$25 m/s}.
\newblock {\em \icarus} 178:253--263.

\bibitem{Fischer.Ciesla.2014} Fischer RA, Ciesla, FJ (2014)
  Dynamics of the terrestrial planets from a large number of N-body
  simulations. Earth and Planetary Science Letters 392: 28--38.


\bibitem{Zsom.etal.2010}
{Zsom} A, {Ormel} CW, {G{\"u}ttler} C, {Blum} J, {Dullemond} CP (2010) {The
  outcome of protoplanetary dust growth: pebbles, boulders, or planetesimals?
  II. Introducing the bouncing barrier}.
\newblock {\em \aap} 513:A57.

\bibitem{Guttler.etal.2010}
{G{\"u}ttler} C, {Blum} J, {Zsom} A, {Ormel} CW, {Dullemond} CP (2010) {The
  outcome of protoplanetary dust growth: pebbles, boulders, or planetesimals?.
  I. Mapping the zoo of laboratory collision experiments}.
\newblock {\em \aap} 513:A56.

\bibitem{Ormel.etal.2008}
{Ormel} CW, {Cuzzi} JN, {Tielens} AGGM (2008) {Co-Accretion of Chondrules and
  Dust in the Solar Nebula}.
\newblock {\em \apj} 679:1588--1610.

\bibitem{Gammie.1996}
Gammie CF (1996) {Layered Accretion in T Tauri Disks}.
\newblock {\em \apj} 457:355--+.

\bibitem{Bai.2014}
{Bai} XN (2014) {Hall-effect-Controlled Gas Dynamics in Protoplanetary Disks.
  I. Wind Solutions at the Inner Disk}.
\newblock {\em \apj} 791:137.

\bibitem{Touboul.etal.2007}
{Touboul} M, {Kleine} T, {Bourdon} B, {Palme} H, {Wieler} R (2007) {Late
  formation and prolonged differentiation of the Moon inferred from W isotopes
  in lunar metals}.
\newblock {\em \nat} 450:1206--1209.

\bibitem{Johansen.etal.2015}
{Johansen} A, {Mac Low} MM, {Lacerda} P, {Bizzarro} M (2015) {Growth of
  asteroids, planetary embryos and Kuiper belt objects by chondrule accretion}.
\newblock {\em ArXiv e-prints}.

\bibitem{Kenyon.Hartmann.1987} Kenyon SJ, Hartmann L (1987)
  Spectral energy distributions of T Tauri stars - Disk flaring and
  limits on accretion.  \newblock {\em \apj} 323:714--733.

\bibitem{Garaud.Lin.2007} Garaud P, Lin DNC (2007) The Effect of
  Internal Dissipation and Surface Irradiation on the Structure of
  Disks and the Location of the Snow Line around Sun-like Stars.
  \newblock {\em \apj} 654:606--624.

\bibitem{Bitsch.etal.2014} Bitsch B, Morbidelli A, Lega E, Crida
  A (2014) Stellar irradiated discs and implications on migration of
  embedded planets. II. Accreting-discs.  \newblock {\em \aap}
  564:A135.

\bibitem{Lecar.Franklin.1973}
{Lecar} M, {Franklin} FA (1973) {On the original distribution of the asteriods.
  I.}
\newblock {\em \icarus} 20:422--436.

\bibitem{Walsh.etal.2012}
{Walsh} KJ, {Morbidelli} A, {Raymond} SN, {O'Brien} DP, {Mandell} AM (2012)
  {Populating the asteroid belt from two parent source regions due to the
  migration of giant planets---''The Grand Tack''}.
\newblock {\em Meteoritics and Planetary Science} 47:1941--1947.

\bibitem{Bottke.etal.2005}
Bottke WF et~al. (2005) {Linking the collisional history of the main asteroid
  belt to its dynamical excitation and depletion}.
\newblock {\em \icarus} 179:63--94.

\bibitem{Marchi.etal.2012}
{Marchi} S et~al. (2012) {The Violent Collisional History of Asteroid 4 Vesta}.
\newblock {\em Science} 336:690--.

\bibitem{Bottke.2014}
{Bottke} WF (2014) {On the Origin and Evolution of Vesta and the V-Type
  Asteroids}.
\newblock {\em LPI Contributions} 1773:2024.

\bibitem{Batalha.etal.2013}
{Batalha} NM et~al. (2013) {Planetary Candidates Observed by Kepler. III.
  Analysis of the First 16 Months of Data}.
\newblock {\em \apjs} 204:24.

\bibitem{Gundlach.Blum.2015}
{Gundlach} B, {Blum} J (2015) {The Stickiness of Micrometer-sized Water-ice
  Particles}.
\newblock {\em \apj} 798:34.

\bibitem{Morbidelli.etal.2009}
Morbidelli A, Bottke WF, Nesvorn\'{y} D, Levison HF (2009) {Asteroids were born
  big}.
\newblock {\em \icarus} 204:558--573.

\end{thebibliography}
\end{document}